# Polarization versus Temperature in Pyridinium Ionic Liquids

Vitaly V. Chaban[1,2] and Oleg V. Prezhdo[1]

[1] MEMPHYS - Center for Biomembrane Physics, Syddansk Universitet, Odense M., 5230, Kingdom of Denmark

[2] Department of Chemistry, University of Southern California, Los Angeles, CA 90089, United States

**Abstract**. Electronic polarization and charge transfer effects play a crucial role in thermodynamic, structural and transport properties of room-temperature ionic liquids (RTILs). These non-additive interactions constitute a useful tool for tuning physical chemical behavior of RTILs. Polarization and charge transfer generally decay as temperature increases, although their presence should be expected over an entire condensed state temperature range. For the first time, we use three popular pyridinium-based RTILs to investigate temperature dependence of electronic polarization in RTILs. Atom-centered density matrix propagation molecular dynamics, supplemented by a weak coupling to an external bath, is used to simulate the temperature impact on system properties. We show that, quite surprisingly, non-additivity in the cation-anion interactions changes negligibly between 300 and 900 K, while the average dipole moment increases due to thermal fluctuations of geometries. Our results contribute to the fundamental understanding of electronic effects in the condensed phase of ionic systems and foster progress in physical chemistry and engineering.

## 1. Introduction

Room-temperature ionic liquids (RTILs)[1-11] constitute an active research field. Applications of RTILs range from high-performance electrolyte solutions[1] to green solvents[12,13] to synthesis and separation setups.[5] Gas capture using ionic liquids stay somewhat apart from the mainstream efforts.[2] RTILs are famous for exhibiting extremely wide liquid temperature ranges, while preserving low volatility even at hundreds of degrees Celsius.[10] In turn, certain RTILs remain liquid well below the normal melting point of ice. It is, therefore, important to understand correlation between molecular level organization of these compounds and their macroscopic properties.

Non-additive interactions, such as electronic polarization of a cation by an anion and charge transfer between these species, influence heavily physical properties of most ionic liquids.[14,15] Polarizable molecular dynamics (MD) simulations emerge in this context as a natural theoretical requirement for a better description of the strong inter-particle binding in these ionic systems.[16-18] Typical RTILs present very slow dynamics, low ionic conductivity, and high viscosity. In addition, structural and dynamical heterogeneities at room temperature distinguish microscopic arrangements of ordinary molecular liquids and RTILs.[10,17]

Although non-additive non-bonding interactions in the condensed state of RTILs have been attended before, their dependence on temperature is unknown. The phenomenon has a crucial impact on the phase diagram of RTILs and numerous phase dependent physical chemical properties. Our work, for the first time, analyzes electronic polarization as a function of temperature for three pyridinium-based RTILs. Quite surprisingly, we find that temperature has little impact on electron delocalization, while it does influence dipole moments of ion pairs. The dipoles grow because of thermal expansion and geometry fluctuations. The study indicates that polarization and charge transfer should be properly represented in simulations of RTILs over the whole temperature range corresponding to the liquid state.

## 2. Simulation Methodology

The electronic energy levels and their populations were obtained for the pyridinium-based ionic liquids, [PY][Cl], [PY][BF$_4$], [PY][N(CN)$_2$], with density functional theory (DFT) using the BLYP functional. BLYP is a well-established, reliable exchange-correlation functional in the generalized gradient approximation. The wave function was expanded using the effective-core LANL2DZ basis set developed by Dunning and coworkers. This basis provides a reasonable tradeoff between accuracy and computational expense, suitable for MD simulations with average-size systems. The wave function convergence criterion was set to $10^{-8}$ Hartree for all calculations. Table 1 presents the basic properties of the systems.

More accurate calculations would require a larger basis set and a hybrid DFT functional. Pure DFT functionals, such as BLYP, tend to overestimate electronic polarization by favoring delocalized electrons. The use of the moderate size basis set counteracts this tendency. Further, our calculations show that temperature has a minor impact on electron delocalization. By overestimating electron delocalization, pure DFT functionals should also overestimate temperature-induced changes in delocalization. Hence, more rigorous and computationally intense calculations with hybrid functionals should confirm our conclusion.

Atom-centered density matrix propagation (ADMP) MD was simulated for 10 ps at 300, 500, and 900 K for the three selected RTIL systems (Table 1). The method computes a classical nuclear trajectory by propagating electron density with auxiliary degrees of freedom. The self-consistent field procedure used every time-step in Born-Oppenheimer MD is avoided. ADMP MD[19] performs somewhat better than Born-Oppenheimer MD in most cases, especially for relatively large systems and complicated convergence cases. It is not necessary to use pseudopotentials on hydrogen atoms or substitute hydrogen atoms by deuterium atoms to increase the nuclear time-step. Preservation of adiabaticity during dynamics should be controlled. In our calculations, the nuclear time-step of 0.1 fs was used at all temperatures. The fictitious electron mass was set to 0.1 a.m.u. Based on extensive preliminary tests, the present setup provides the best balance between electron-nuclear adiabaticity and fast propagation of the equations-of-motion. The functions of core electrons carried larger weights than those of valence electrons.

Each system was equilibrated for 1 ps to the desired temperatures by velocity rescaling. Next, 10 ps production trajectories were generated. In principle, thermostatting can be turned off after equilibration, and the subsequent simulation can be carried out in the microcanonical ensemble. Due to the small system size, and consequently, large temperature fluctuation, we compared the real and target temperatures every 5.0 fs during the production runs, and rescaled the velocities if the difference exceeded 50 K. The three systems were studied at the three different temperatures (Table 1), resulting in 90 ps of equilibrium trajectory data and 900 000 time-steps of nuclear motion. Sampling of rare conformations of the RTIL ion pairs extends beyond our simulation schedule.

The atomic charges and ion pair dipole moments were generated along the MD trajectories using the Hirshfeld scheme.[20] The electrostatic potentials (ESP) were obtained for optimized system geometries. The ESP derived from the electronic structure were reproduced using a set of point charges centered on each atom, including hydrogen atoms. The atom spheres were defined according to the CHELPG scheme.[21] Figure 1 depicts the optimized geometries (0 K) of isolated ions and ion pairs. The atoms are colored according to their polarity.

The electronic structure computations were performed using GAUSSIAN 09, revision D. The subsequent analysis was performed using utilities developed by V.V.C.

**Table 1**. Simulated systems and selected results

| System | # electrons | # basis functions | Reference temperature, K | Dipole moment, D | Charge on cation, e |
|---|---|---|---|---|---|
| [PY][Cl] | 58 | 393 | 300 | 8.6 | 0.52 |
| [PY][Cl] | 58 | 393 | 500 | 9.7 | 0.54 |
| [PY][Cl] | 58 | 393 | 900 | 10.8 | 0.54 |
| [PY][BF$_4$] | 92 | 580 | 300 | 12.5 | 0.82 |
| [PY][BF$_4$] | 92 | 580 | 500 | 13.2 | 0.82 |
| [PY][BF$_4$] | 92 | 580 | 900 | 16.5 | 0.80 |
| [PY][N(CN)$_2$] | 84 | 580 | 300 | 11.2 | 0.74 |
| [PY][N(CN)$_2$] | 84 | 580 | 500 | 11.8 | 0.71 |
| [PY][N(CN)$_2$] | 84 | 580 | 900 | 13.0 | 0.71 |

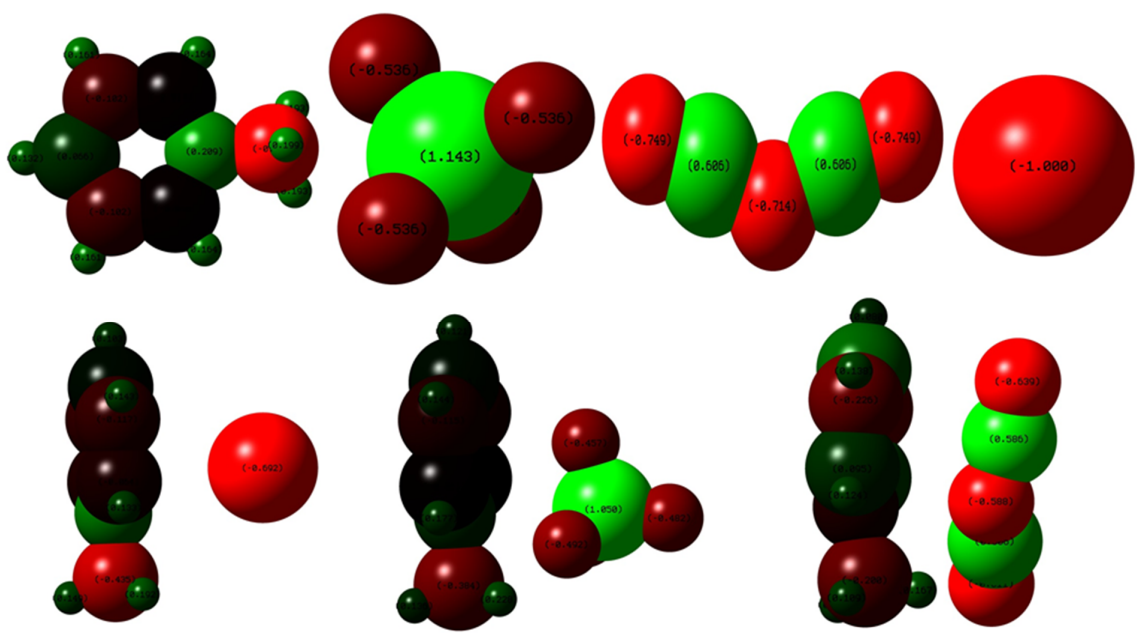

**Figure 1**. Distribution of partial electrostatic charges over the pyridinium cation and the selected anions. The depicted charges have been computed using the CHELPG fitting scheme.[21] The fitting was constrained to reproduce the dipole moments computed from the DFT electron density. Electron-rich atoms are colored in red, electron-poor atoms are colored in green, and neutral atoms are colored in grey/black. The colors are more intense for larger partial charges.

## 3. Results and Discussion

The anions are chosen to represent a variety of structures. Cl⁻ has high charge density and is capable of significantly polarizing surrounding species. Ionic liquids containing Cl⁻ are strongly electrostatically driven and, therefore, exhibit relatively high melting points. Similarly to Cl⁻, $BF_4^-$ is a spherical particle, but with a different distribution of electrons. In turn, $N(CN)_2^-$ features a unique structure with two chemically non-equivalent nitrogen atoms and a symmetry, which may or may not be broken in the condensed phase. The anions with more delocalized excess electronic charge are beyond the scope of the present work, since their polarizing ability is generally weaker. The cation is represented by methylpyridinium, $PY^+$. Pyridinium cations used in practice possess longer hydrocarbon chains (for instance, N-butylpyridinium). Our choice is made in the view of computational limitations of DFT. Electrically neutral hydrocarbon chains neither polarize the central fragment of the cation, nor get polarized by the anion. Pyridinium-based ionic liquids are vigorously investigated nowadays due to their versatility and tunable properties.[22-26]

Table 1 introduces the simulated systems, and their most important parameters and properties. Interestingly, the methyl carbon atom is the most electron deficient atom in the isolated methylpyridinium cation (-0.4e), while all other cation atoms are essentially neutral. The nitrogen atom is only slightly positive, +0.2e. Therefore, one expects that the anion species are coordinated by these two bonded atoms. The boron atom of the tetrafluroborate anion is +1.143e, compensated by the fluorine atoms, -0.536e each. Note that the computed electrostatic point charges at $BF_4^-$ are in excellent agreement with those used in the classical force fields to reproduce electrostatic non-bonded interactions. This fact indicates that our electronic structure calculations give reliable results. The central nitrogen atom of $N(CN)_2^-$ is not symmetrically equivalent to the nitrogen atoms of the two cyan groups. It should be stressed that all three nitrogen atoms carry anion's negative charge, whereas the two carbon atoms are positive, partially compensating the negative charge. Since all three nitrogens are negative, the excess electron is delocalized, in contrast to chloride. Based on the difference in electronegativity between carbon and nitrogen, one expected a smaller difference in their partial electrostatic charges. As expected, coordination of all anions occurs through the nitrogen atom of the pyridine ring and the methyl group. If longer alkyl chains had been used, the positive charge would not have been localized so much on the methyl carbon atom, and the coordination center would have shifted towards the nitrogen atom.

Partial insights into electron polarization and transfer can be obtained from localization of valence orbitals, in particular the highest occupied (HOMO) and lowest unoccupied (LUMO) molecular orbitals of the ion pair (Figure 2). Both HOMO and LUMO are delocalized over the in [PY][Cl] ions. In the case of [PY][BF$_4$], HOMO is localized on the anion, while LUMO is on the cation. Excitation from HOMO to LUMO results in electron transfer from the anion to the cation. HOMO and LUMO of [PY][N(CN)$_2$] are mostly on the anion, but they are not perfectly localized. The analysis of higher-energy orbitals of the ion pairs suggests a strong non-covalent interaction (mostly electrostatic attraction) with evident non-additive phenomena. Note that polarization effects can be viewed as partial excitation of electron from occupied to vacant orbitals.

Further insights can be derived by comparing the total electronic density in the ion pair to the total electronic density of lone ions. Ghost basis functions must be present to avoid artifacts. These possible artifacts are connected with basis set superposition, upon consideration of lone ions in the same conformations as they are in the optimized geometry of the ion pair. Note that the results of Figures 1 and 2 correspond to the electronic structures of cold (0 K) conformations, obtained after geometry optimization.

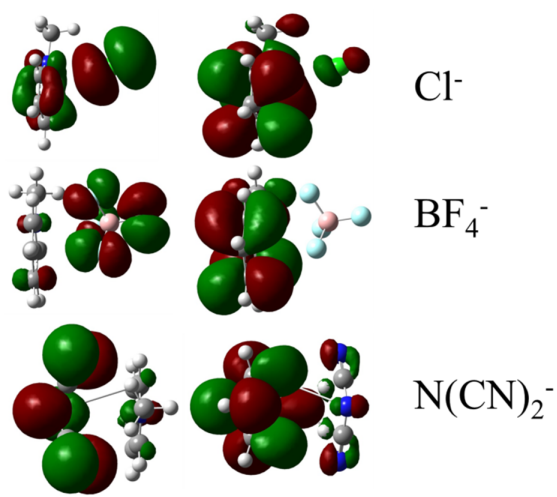

**Figure 2**. Highest occupied molecular orbitals (HOMO, left) and lowest unoccupied molecular orbitals (LUMO, right) in methylpyridinium chloride (Cl⁻), tetrafluoborate (BF$_4^-$), and dicyanamide (N(CN)$_2^-$).

Thermal fluctuations of the total electrostatic partial charge localized on the pyridine ring are plotted in Figure 3 for the three RTILs at the three studied temperature. Huge fluctuations are observed, with fluctuation amplitude in direct proportion to temperature. For instance at 900 K, the sum of Hirshfeld charges[20] on the cation in [PY][Cl] drops down to nearly 0.2e, but then quickly returns to the equilibrium value of 0.54e. No drift in time is observed for this property, suggesting that all systems are properly equilibrated. Surprisingly, the average values are almost independent of temperature (Table 1, Figure 4). It is unexpected, since polarization tends to decay as temperature increases. A 600 K temperature increase from 300 to 900 K is quite significant to expect changes. To exclude the possibility that the lack of the temperature impact arises because the simulated systems are small, leading to a large uncertainty in the actual temperature, we performed additional simulations using four ion pairs. However, the conclusion did not change. The canonically averaged charges are nearly independent of temperature; however, the magnitude of the fluctuations around the average value is proportional to temperature.

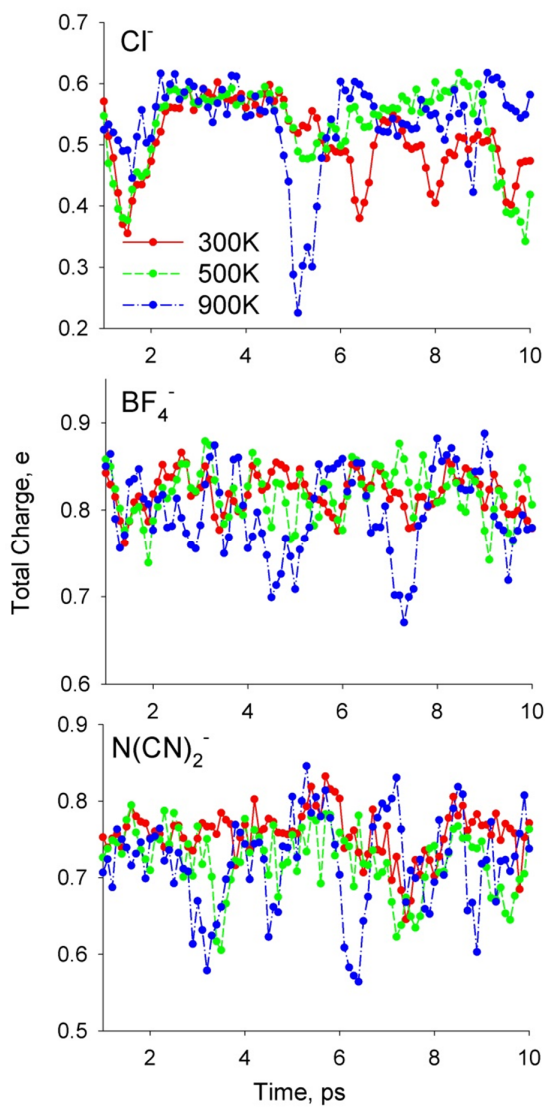

**Figure 3**. Fluctuation in the positive charge localized on the pyridine ring of the cation in the presence of the Cl$^-$, BF$_4^-$, and N(CN)$_2^-$ anions: 300 K (red solid line), 500 K (green dashed line), and 900 K (blue dash-dotted line). Electron partitioning between the atoms is based on the Hirshfeld procedure.[20]

The temperature-induced change in the negative charge of the chloride anion is within the 0.52-0.54e range. The changes in the negative charges of BF$_4^-$ and N(CN)$_2^-$ are 0.80-0.82e and 0.71-0.74e, respectively. We hypothesize that insensitivity to temperature comes from genuinely strong non-covalent interactions in these systems. The cation-anion binding energy well exceeds the kinetic energy in the binding coordinate. All ions remained stable during 10 ps of the simulated nuclear dynamics, but this time period may be too short to observe a decomposition reaction.

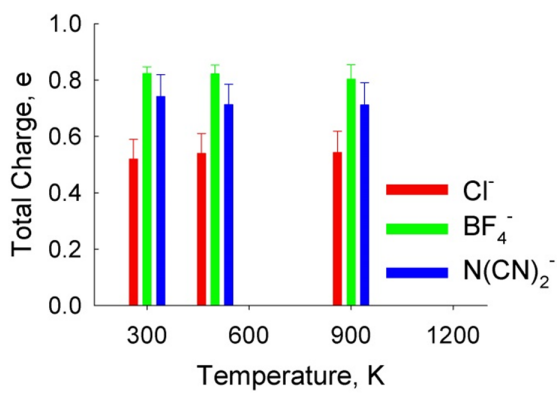

**Figure 4**. Canonically averaged values of the positive charge localized on the pyridine ring of the cation for different temperatures and anions. Electron partitioning between the atoms is based on the Hirshfeld procedure.[20]

Dipole moment (Figure 5) provides an alternative description of electric properties, as it characterizes not only atomic charges, but also separation of positive and negative centers-of-mass of the electron-nuclear system. Dipole moments of the ion pairs of the pyridinium-based RTILs increase significantly with temperature (Table 1). The largest dipole moment of 16.5 D is observed for [PY][$BF_4$] at 900 K. The dipole moments of [PY][Cl] are systematically smaller at all temperatures: 8.6 D (300 K), 9.7 D (500 K), 10.8 D (900 K). The observed growth of dipole moments with temperature leads to increase in the dielectric constant of these RTILs in the condensed phase.

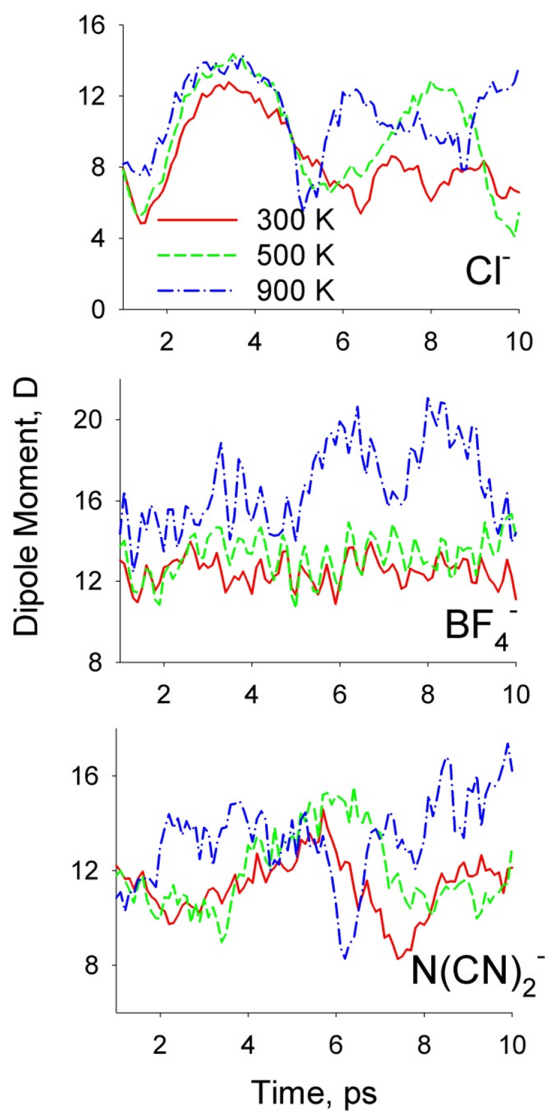

**Figure 5**. Evolution of the dipole moment of the ion pair containing pyridinium cation at various temperatures: 300 K (red solid line), 500 K (green dashed line), and 900 K (blue dash-dotted line).

## 4. Concluding Remarks

We have shown that temperature has a very minor impact on electron delocalization in the pyridinium-based RTIL ion pairs, which is a rather unexpected result. At the same time, dipole moments increase systematically upon RTIL heating due to growing thermal fluctuations of the corresponding structures, and thermal expansion. Since electronic polarization and charge transfer effects do not decay with temperature, they must be properly accounted for in simulations of condensed phase ionic liquids.

The provided analysis is based on ion pairs simulated in vacuum rather than condensed phase. This is due to a set of methodological restrictions. First, computational cost for simulations of quantum dynamics is often prohibitive for systems of more than a few hundred electrons. Second, computation of the long-range electrostatic potential is not straightforward in periodic systems. In addition, point charges cannot be univocally defined far from the surface of an ion cluster/aggregate ("buried atom" problem). Third, the dipole moment is most meaningful in the case of a single ion pair, whereas higher-order electric moments suit better for description of large molecular formations. We expect that periodic systems would have slightly different electron densities on each atom from those computed here for the ion pairs. Nevertheless, the qualitative conclusions of this work should not change. In the condensed phase, ion pairs would dominate over more complicated ionic aggregates. Such systems exhibit a well-defined first peak in the radial distribution functions, but much less structured second and possibly third peaks. If a single cation-anion coordination site prevails, as it is anticipated in this case, introduction of additional ions will not change the observables qualitatively.

**Acknowledgments**

This research has been supported by grant CHE-1300118 from the US National Science Foundation. MEMPHYS is the Danish National Center of Excellence for Biomembrane Physics. The Center is supported by the Danish National Research Foundation.

**Author Information**

E-mail address for correspondence: vvchaban@gmail.com; chaban@sdu.dk (V.V.C.)

**REFERENCES**

(1) Fedorov, M. V.; Kornyshev, A. A. *Chemical Reviews* **2014**, *114*, 2978.
(2) Brennecke, J. E.; Gurkan, B. E. *Journal of Physical Chemistry Letters* **2010**, *1*, 3459.
(3) Lei, Z. G.; Dai, C. N.; Chen, B. H. *Chemical Reviews* **2014**, *114*, 1289.
(4) Ren, Z.; Ivanova, A. S.; Couchot-Vore, D.; Garrett-Roe, S. *Journal of Physical Chemistry Letters* **2014**, *5*, 1541.
(5) Hallett, J. P.; Welton, T. *Chemical Reviews* **2011**, *111*, 3508.
(6) van Rantwijk, F.; Sheldon, R. A. *Chemical Reviews* **2007**, *107*, 2757.
(7) Luo, X. Y.; Ding, F.; Lin, W. J.; Qi, Y. Q.; Li, H. R.; Wang, C. M. *Journal of Physical Chemistry Letters* **2014**, *5*, 381.
(8) Zhang, X. X.; Liang, M.; Ernsting, N. P.; Maroncelli, M. *Journal of Physical Chemistry Letters* **2013**, *4*, 1205.
(9) Hettige, J. J.; Kashyap, H. K.; Annapureddy, H. V. R.; Margulis, C. J. *Journal of Physical Chemistry Letters* **2013**, *4*, 105.
(10) Chaban, V. V.; Prezhdo, O. V. *Journal of Physical Chemistry Letters* **2013**, *4*, 1423.
(11) Szefczyk, B.; Cordeiro, M. N. D. S. *Journal of Physical Chemistry B* **2011**, *115*, 3013.
(12) Horvath, I. T.; Anastas, P. T. *Chemical Reviews* **2007**, *107*, 2169.
(13) Maciel, C.; Fileti, E. E. *Chemical Physics Letters* **2013**, *568*, 75.
(14) Chaban, V. *Physical Chemistry Chemical Physics* **2011**, *13*, 16055.
(15) Chaban, V. V.; Voroshylova, I. V.; Kalugin, O. N. *Physical Chemistry Chemical Physics* **2011**, *13*, 7910.
(16) Borodin, O. *Journal of Physical Chemistry B* **2009**, *113*, 11463.
(17) Lopes, J. N. C.; Padua, A. A. H. *Theoretical Chemistry Accounts* **2012**, *131*.
(18) Salanne, M.; Rotenberg, B.; Jahn, S.; Vuilleumier, R.; Simon, C.; Madden, P. A. *Theoretical Chemistry Accounts* **2012**, *131*.
(19) Schlegel, H. B.; Iyengar, S. S.; Li, X. S.; Millam, J. M.; Voth, G. A.; Scuseria, G. E.; Frisch, M. J. *Journal of Chemical Physics* **2002**, *117*, 8694.
(20) Hirshfeld, F. L. *Theoretica Chimica Acta* **1977**, *44*, 129.
(21) Breneman, C. M.; Wiberg, K. B. *Journal of Computational Chemistry* **1990**, *11*, 361.
(22) Kim, M. J.; Shin, S. H.; Kim, Y. J.; Cheong, M.; Lee, J. S.; Kim, H. S. *Journal of Physical Chemistry B* **2013**, *117*, 14827.
(23) Oliveira, M. B.; Llovell, F.; Coutinho, J. A. P.; Vega, L. F. *Journal of Physical Chemistry B* **2012**, *116*, 9089.
(24) Aparicio, S.; Atilhan, M. *Journal of Physical Chemistry B* **2012**, *116*, 2526.
(25) Bandres, I.; Alcalde, R.; Lafuente, C.; Atilhan, M.; Aparicio, S. *Journal of Physical Chemistry B* **2011**, *115*, 12499.
(26) Fernandes, A. M.; Rocha, M. A. A.; Freire, M. G.; Marrucho, I. M.; Coutinho, J. A. P.; Santos, L. M. N. B. F. *Journal of Physical Chemistry B* **2011**, *115*, 4033.

TOC Graphic

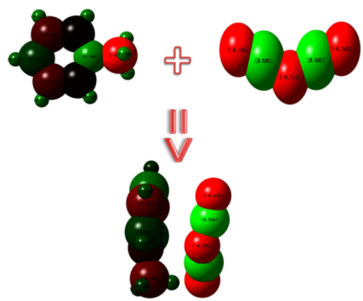